# Bandgap and doping effects in $MoS_2$ measured by Scanning Tunneling Microscopy and Spectroscopy

*Chih-Pin Lu[1,2], Guohong Li[1], Jinhai Mao[1], Li-Min Wang[2] and Eva Y. Andrei[1]*

[1] Department of Physics and Astronomy, Rutgers University, Piscataway, New Jersey 08855, USA

[2] Department of Physics, National Taiwan University, Taipei 10617, Taiwan

Abstract

The discovery of graphene has put the spotlight on other layered materials including transition metal dichalcogenites (TMD) as building blocks for novel heterostructures assembled from stacked atomic layers. Molybdenum disulfide, $MoS_2$, a semiconductor in the TMD family, with its remarkable thermal and chemical stability and high mobility, has emerged as a promising candidate for post-silicon applications such as switching, photonics, and flexible electronics. Since these rely on controlling the position of the Fermi energy ($E_F$), it is crucial to understand its dependence on doping and gating. Here we employed scanning tunneling microscopy (STM) and spectroscopy (STS) with gating capabilities to measure the bandgap and the position of $E_F$ in $MoS_2$, and to track its evolution with gate voltage. For bulk samples, the measured bandgap (~1.3eV) is comparable to the value obtained by photoluminescence, and the position of $E_F$ (~0.35eV) below the conduction band, is consistent with n-doping reported in this material. Using topography together with spectroscopy we traced the source of the n-doping in bulk $MoS_2$ samples to point defects, which we attribute to S vacancies. In contrast, for thin films deposited on $SiO_2$, we found significantly higher levels of n-doping that cannot be attributed to S vacancies. By combining gated STS with transport measurements in a field effect transistor (FET) configuration, we demonstrate that the higher levels of n-doping in thin film samples is due to charge traps at the sample-substrate interface.

The development of techniques to isolate atomic layers[1] and to integrate them into atomically precise heterostructures[2] has enabled the design of novel material properties and applications by exploiting the proximity between layers with different electronic structures such as metallic (Graphene), insulating (hBN) and semiconducting (2H-MoS$_2$) [3-6]. 2H-MoS$_2$, a layered material in the transition metal dichalcogenite family has recently attracted much attention owing to its remarkable electrical and optical properties. The bulk MoS$_2$ has an indirect bandgap of 1.2~1.3 eV[7-10] which, due to quantum confinement, crosses over to a direct bandgap of ~1.9 eV when the material is exfoliated down to a monolayer[11]. Thin layers of MoS$_2$ are well suited as a channel material in FET applications exhibiting high room temperature mobility, almost ideal switching characteristics, and low standby power dissipation[6, 12-14]. Furthermore, the absence of dangling bonds and of surface states and its resistance to oxidation make MoS$_2$ a particularly good candidate for STM studies. In both bulk and thin layers of MoS$_2$ deposited on SiO$_2$ the conductivity consistently exhibits n-type character[6, 15-17], but to date source of the doping remains unclear. Proposals include substitutional impurities[14, 18] such as Cl, Br or Re[19], S vacancies[20-22], and impurities trapped at the interface with the SiO$_2$ substrate[3, 23]. Here we employ STM/STS combined with transport measurements to elucidate the nature of doping in MoS$_2$ and its connection to the switching characteristics of the thin layers.

Samples were prepared by exfoliating bulk *2H*-MoS$_2$ crystals (purchased from SPI) onto SiO$_2$ (300nm)/Si substrates to achieve the desired sample thickness and to expose a fresh sample surface. MoS$_2$ flakes were identified by optical microscopy and subsequently characterized by atomic force microscopy (AFM) and Raman spectroscopy in ambient conditions. The AFM height profile of one of the samples, shown in Fig.1(a), indicates a thickness of 4.2 nm corresponding to 6 layers. The

Raman spectra, Fig.1(b), for single, multilayer and bulk MoS$_2$ samples show a monotonic evolution with the number of layers[24]. For the 6-layer sample the spectrum is already very close to that of bulk. FET devices were fabricated using e-beam lithography with Ti/Au (2 nm/60 nm) contacts deposited by electron-beam evaporation at a base pressure of 2 × 10$^{-7}$ Torr. Prior to measurements the MoS$_2$ devices were baked for 3hrs in forming gas (90% Ar, 10%H$_2$) at 230 $^0$C. STM and STS measurements were performed at 80 K in a home-built STM[25-27] using mechanically cut Pt-Ir tips. The STM images were recorded in constant current mode with the bias voltage, $V_b$, applied to the sample while the STM tip was held at ground potential. The differential conductance (dI/dV) spectra were measured using a lock-in technique (ac modulation: 5mV$_{rms}$ at 440Hz) with fixed tip to sample distance. Gated STS measurements were carried out in a device configuration, shown in Fig.1(c), in which a thin sample was deposited on a 300nm SiO$_2$ substrate capping a degenerately p-doped Si gate.

To investigate the source of doping in thin MoS$_2$ samples we employed a 6 layer sample configured in an FET device which combines STS with gating capability. As shown in the STM topography image in Fig.2(a) the sample surface is relatively flat and clean with corrugations not exceeding ~ 0.2nm over a 50nm × 50nm area. Zooming in to atomic resolution in Fig.2(b) reveals the sulfur surface atoms arranged in a triangular lattice with lattice constant ~ 0.31 nm, in agreement with the accepted crystallographic value, of 0.316 nm. This indicates high surface quality with no surface contamination left over from the fabrication process. STS measurements at the same position revealed a bandgap of 1.2 ± 0.1 eV that is comparable to the reported photoluminescence gap[10], suggesting that tip induced band-bending does not significantly affect our measurements[28]. Using the setup shown in Fig.1(c) we studied the local electronic properties as a function of gate voltage ($V_g$) across the SiO$_2$ substrate. For $V_g = 0$V, the position of $E_F$ at the edge of the conduction band (CB) indicates the presence of shallow donors, which is in agreement with previous reports of unintentional n-doping of such devices[3, 6]. A positive $V_g$ slowly pushes $E_F$ into CB (Fig.2(d)), reflecting the high density of states near the bottom of CB. In contrast, a negative $V_g$ moves $E_F$ away from the CB, making the sample less conductive. The effect of negative gating suddenly accelerates at a threshhold voltage $V_{gt} \sim -15$V as shown in Fig.2(d). Applying $V_g = -25$V shifts the position of $E_F$ to the center of the bandgap, as shown in Fig.2(c). In this regime special care had to be taken to avoid crashing the tip and damaging sample, as described in the supplementary material (S1) the sample, as described in the supplementary material (S1).

From the evolution of $E_F$ with $V_g$ in the gap region, inset in Fig.2(d), we estimate $dE_F/dV_g = 0.06q$, where $q$ is the magnitude of the electron charge. Rewriting this expression in terms of the quantum capacitance[29], $C_Q = q^2 \frac{dn}{dE_F}$, where $n$ is the carrier density per unit area, we obtain $\frac{dE_F}{dV_g} = \frac{1}{q}\frac{dE_F}{dn}\frac{dqn}{dV_g} = q\frac{C}{C_Q}$. Noting that $\frac{dqn}{dV_g} = C = (1/C_{ox} + 1/C_Q)^{-1}$ is the total device capacitance resulting from the series connection of $C_Q$ with the oxide capacitance, $C_{ox} = \varepsilon_{ox}/d_{ox} \approx 12 nF \cdot cm^{-2}$, and rewriting $\frac{dE_F}{dV_g} = 0.06q = q\frac{C}{C_Q}$, we find $C_Q/C_{ox} = 15.6$. This gives $C_Q = 187 nF \cdot cm^{-2}$, $C = 11.3 nF \cdot cm^{-2}$ for our device. From the quantum capacitance we can obtain the total density of states per unit area: $\frac{dn}{dE_F} \equiv D(E_F) \sim 1.16 \times 10^{12}$ states/eV cm$^{-2}$. We next compare the STS results to transport measurements of a thin sample in an FET device configuration.

In order to correlate the STS data with transport measurements we used a 2-terminal device configuration shown in Fig.3 to follow the gate dependence of the drain source current, $I_{ds}$ $(V_g)$ in 4 samples (Supplementary material S2). We found that the device characteristics such as mobility and subthreshold slope improve substantially after *in-situ* annealing in vacuum ($10^{-6}$ mbar) at 120 $^0$C for 20 hours, similar to results reported in the literature[3]. For the bilayer sample shown in Fig.3(b), the conduction threshold, initially $V_{gt}$ = 8V, decreased after annealing to $V_{gt}$ = -10V approaching the value obtained by STS. From the measured gate dependence in the linear regime of $I_{ds}$ $(V_g)$ we obtain the field-effect mobility at 300K $\mu = \frac{L}{WC_{ox}} \frac{1}{V_{ds}} \frac{dI_{ds}}{dV_g}$ ~3cm$^2$ V$^{-1}$ s$^{-1}$ which increased to ~6 cm$^2$ V$^{-1}$ s$^{-1}$ after annealing. Here L = 7.4 μm, W = 6.5 μm are the channel length and width, respectively. The observed monotonic increase of the post-anneal mobility with number of layers (Supplementary material S2) indicates that scattering is dominated by surface defects. Another important metric of device quality is the subthreshold swing[30] defined as S = $(ln10)k_BT/q$ $(1+C_Q/C_{ox})$. The subthreshold swing, which can also be expressed as $S = \frac{dV_g}{d(log\, I_{ds})}$, characterizes the switching time of the device: the lower S the faster the device. Its value can be obtained from $I_{ds}$ $(V_g)$ by measuring the subthreshold slope, $\frac{d(log\, I_{ds})}{dV_g}$, in the exponential regime of the curve, below the threshold voltage ($V_g < V_{gt}$). For $C_Q \ll C_{ox}$ the device reaches its thermodynamically limited fastest performance with $S_{min}$ = $(ln10)k_BT/q$ = 60mV/decade at T = 300K. This is significantly below the S values measured in our devices indicating that they operate in the opposite limit $C_Q \gg C_{ox}$. For the sample discussed here, S = 1.5V/decade at T=300K prior to annealing, gives an estimated $C_Q$ ~ 288nF cm$^{-2}$ $\gg C_{ox}$. The device improved after annealing dropping to S = 0.7V/decade, and $C_Q$ ~ 128nF cm$^{-2}$. Remarkably, the values of $C_Q$ obtained by transport agree with the STS values within the experimental uncertainty, indicating that transport measurements can provide a reliable measure of the quantum capacitance.

The weak thickness dependence of S together with the monotonic thickness dependence of the post-anneal mobility (Supplementary material S2), suggests that surface traps at the $SiO_2$ interface significantly affect the device performance. In order to trace the origin of the doping and to establish whether it is due to bulk or surface traps, we carried out STM/STS on a freshly cleaved bulk sample whose surface has not undergone wet treatment and that is not supported by a $SiO_2$ substrate. The sample was mounted into the STM head using silver paint to attach the back of the sample to a reference electrode. STS measurements on the bulk sample reveal a band gap of ~ 1.3 eV, Fig.4(c), consistent with the photoluminescence results. Contrary to the results in the thin sample, $E_F$ is now ~ 0.35 eV below the CB indicating the presence of deep donors. Indeed large area (50 nm×50 nm) STM topography, shown in Fig. 4(a,b), which was obtained at $V_b$ = 1.2V reveals defects as dark spots. Three types of defects can be distinguished by their apparent depth and diameter: darkest (-0.12 nm in depth and 3 nm in diameter); intermediate (-0.05 nm and 4 nm); faint (-0.02 nm and 5.5 nm). Zooming into the defects to resolve the atomic structure we find that the lattice periodicity is intact suggesting that the defects are buried under the surface (Supplementary material S3, S4). According to *ab initio* calculations, the electronic structure of point defects on the surface of $MoS_2$ are tightly localized within one atomic distance[19]. By contrast, the defects observed here are extended several lattice spacings without a localized feature on the central atom. This again indicates defects buried under the surface, similar to the case of buried Mn impurities in GaAs or InAs which produce a topography signature extending over several lattice spacing[31-33]. Amongst the reported substitution atoms in $MoS_2$, only Cl, Br and Re are consistent with n-doping. However the ionization energy of these donors, ~0.16 eV, 0.26 eV and

0.06 eV respectively[18, 19], is too low to account for the observed $E_F$ ~ 0.35 eV below the CB. Furthermore, X-Ray Fluorescence (XRF) and X-ray photoelectron spectroscopy (XPS) analysis of our crystal indicated that within the experimental resolution of 100 ppm these elements were not present. As we show next the observed defects are consistent with S vacancies.

Zooming into the center of a defect, Fig. 4(d), the STS spectra reveal a sharp in-gap peak at $V_b = -0.94V$ that is close to the valence band and not near the CB edge where donor states are expected. This apparent discrepancy reflects the tip-induced band bending[28]. When the bias voltage corresponds to the flat band condition the donor state crosses $E_F$ creating a new conduction path where donor electrons tunnel into empty states of the STM tip which gives rise to a peak in the differential conductance. We can calculate the donor ionization energy, $E_d$, from the flat band condition: $qV_{FB} = E_A + \frac{E_d}{2} - \Phi_{tip}$, where $V_{FB}$ = -0.94 eV the flat-band bias voltage, $E_A$ = 4.2 eV the electron affinity in $MoS_2$[34, 35], and $\Phi_{tip}$ the work function of the tip. Using the value[36] $\Phi_{tip}$ = 5.5eV for the Pt-Ir tip, we obtain $E_d$ ~ 0.7 eV below the CB consistent with the position of $E_F$ and with the calculated values for S vacancies[18-22]. This gives an estimate of the defect size[32]: $R \sim \frac{\hbar}{\sqrt{2m^*E_d}} = 0.3 nm$ (m* = 0.4$m_e$ is the effective electron mass) which agrees with the calculated size of an S vacancy[19]. However, when buried under the surface, point defects produce a more extended topography image which typically increases with the depth[31-33], consistent with the fact that the atomic topography image shows no missing surface atoms (Fig. 4(b)). The three defect sizes can be understood in terms of vacancies located in different S layers underneath the surface [32, 33]. To estimate the density of vacancies we measure the surface density of defects of equal size, $N_l \approx 3.5 \times 10^{10} cm^{-2}$ and assuming the same defect density in all S layers we obtain the volume density:

$N_{3d} = N_l / a \approx 1.1 \times 10^{18} cm^{-3}$ where $a$ = 0.31 nm is the vertical separation between sulfur layers (Supplementary material S5). At room temperature this gives an activated bulk carrier density of $n \sim N_{3d} e^{-(E_F - E_D)/k_B T} \approx 10^{13} cm^{-3}$ in the ungated devices. We can now estimate the equivalent areal density of activated carriers in the double layer, t ~ 2.5 nm, used in the FET device above to be $n_0 = nt \sim 2.5 \times 10^4 cm^{-2}$, which is significantly lower than the measured residual density in thin devices. These results clearly exclude bulk mechanisms such as substitutional dopants or sulfur vacancies, as the source of the post-anneal doping observed in thin layer MoS$_2$. We must conclude that the n-doping in the thin layer is due to trapped charges at the interface with the SiO$_2$ substrate, consistent with the observed variability in device quality, with the post-anneal decrease in the subthreshold swing[3, 23], and with the monotonic thickness dependence of the post-anneal mobility. Likely candidates for the trapped charges could be Na ions which are notoriously prone to contaminate[37] the surface of SiO$_2$. For example in the case of graphene deposited on SiO$_2$ it is well known that trapped Na ions at the interface cause significant levels of unintentional doping, however there Na produces holes whereas in MoS$_2$ it induces electron doping[23, 38].

In summary by combining STM, STS and gating capabilities together with transport measurements we have traced the intrinsic n-doping in bulk 2H-MoS$_2$ to point defects which are consistent with S vacancies. Within the experimental error we found no evidence of Cl, Br, or Re dopants. Furthermore, we demonstrated that the significantly higher n-doping observed in thin films deposited on SiO$_2$ is extrinsic and can be attributed to trapped donors at the interface with the SiO$_2$ substrate.

**Supporting Information Available:**

**Acknowledgements**

Funding was provided by DOE-FG02-99ER45742 (E.Y.A, G.L, J.M), NSF DMR 1207108 (E.Y.A., C.P.L) and National Science Council of Taiwan (C.P.L)

**Author information**

Correspondence and requests for materials should be addressed to Eva. Y. Andrei (eandrei@physics.rutgers.edu).

**FIGURES:**

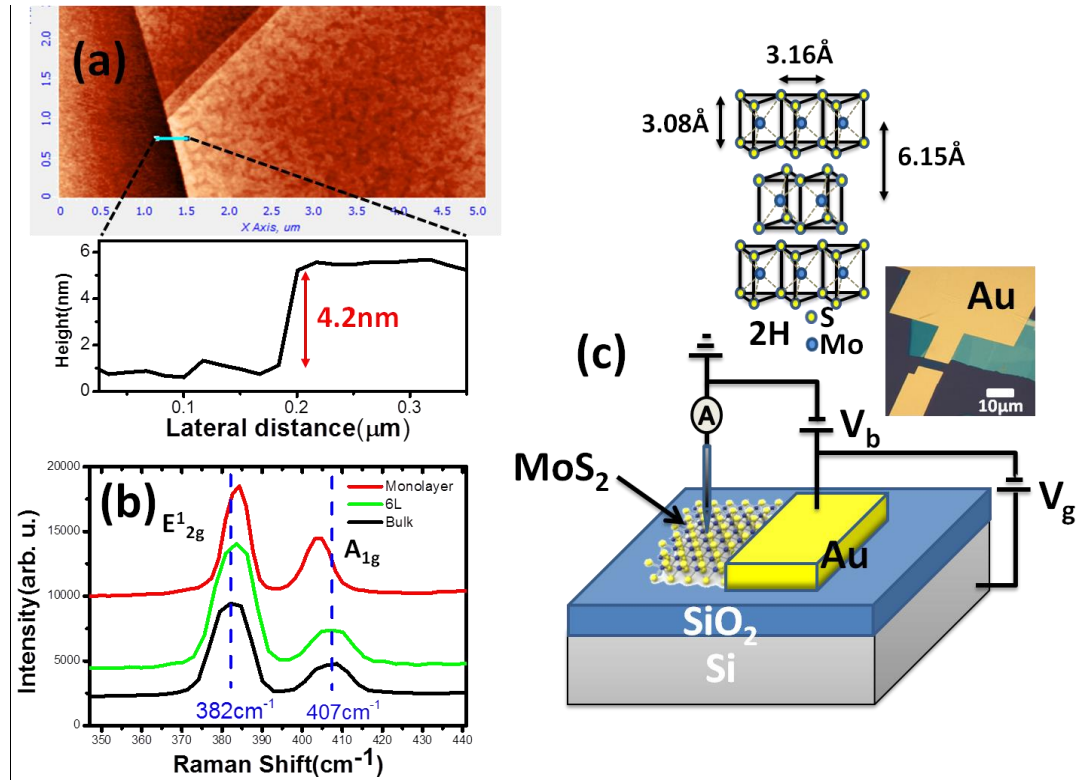

**Figure 1.** (a) AFM image of multilayer $MoS_2$ on a $SiO_2$/Si substrate. The step height indicates a thickness of 4.2nm corresponding to 6 layers. (b) Raman spectrum showing the in-plane $E^1_{2g}$ and out-of-plane $A_{1g}$ resonances of single, multilayer and bulk $MoS_2$. The laser wavelength was 633 nm and the spot size ~2 μm. (c) Schematic of $MoS_2$ sample mounted in an STM configuration with gating capabilities. The sample bias $V_b$ is applied between the STM tip and the $MoS_2$ sample contacted by a Ti/Au electrode. The back gate voltage $V_g$ is applied between the p-doped Si substrate and the top electrode. Inset: Optical micrograph of a mechanically exfoliated multilayer $MoS_2$ flake and schematics of the atomic structure of 2H-$MoS_2$.

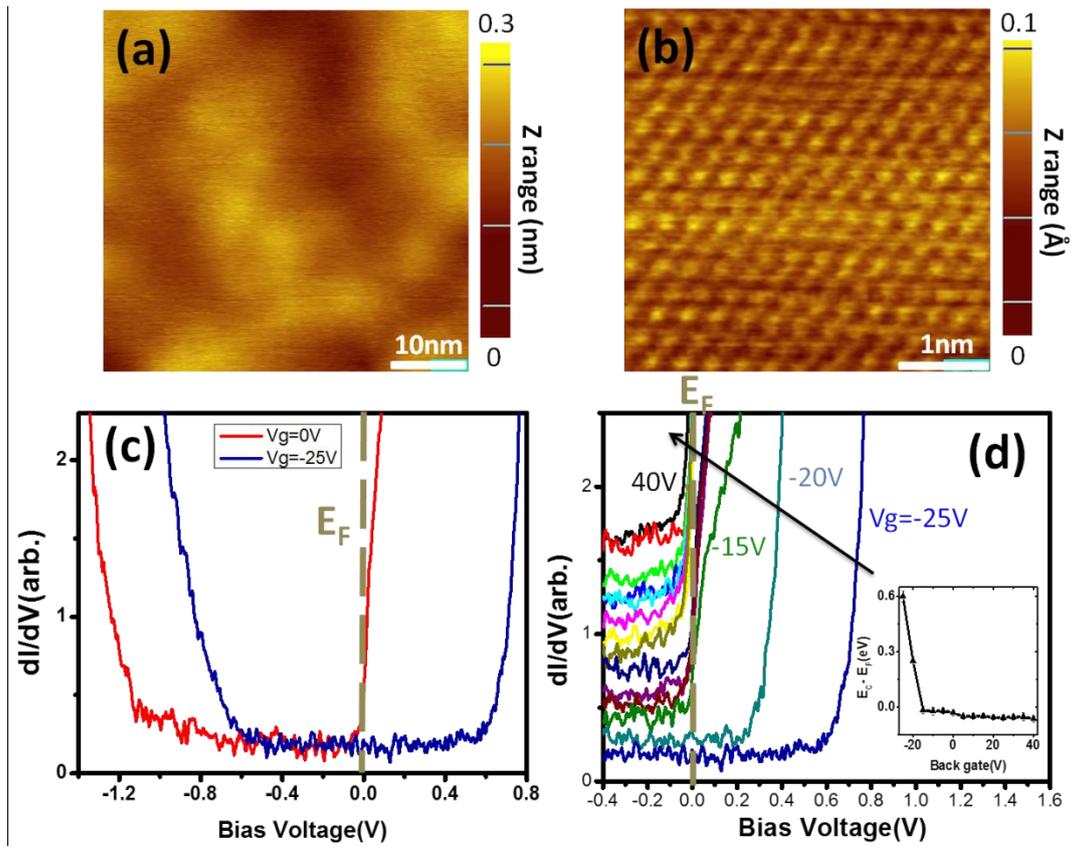

**Figure 2.** (a) STM constant current topography image of a 6 layer $MoS_2$ sample measured in the device configuration of Fig. 1(c) for $V_b = 1.4$ V and $I = 20$ pA. (b) Same as (a) zooming in to atomic resolution. (c) dI/dV spectrum taken in the area shown in (b) for $V_g = 0$V (red) and $V_g = -25$V (blue) (d) Evolution of dI/dV spectrum with $V_g$. Curves are offset vertically and only show the CB edge. Inset: Gate voltage dependence of the position of the Fermi energy relative to the bottom of the conduction band.

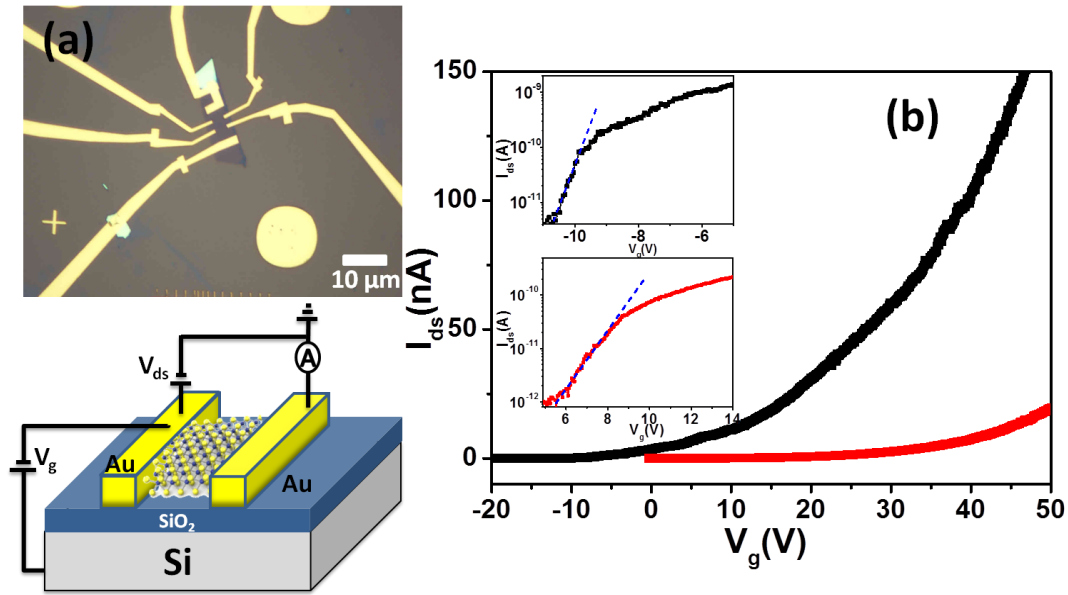

**Figure 3.** (a) Top: Optical micrograph of a thin $MoS_2$ device supported on a 300nm-thick $SiO_2$ substrate configured for a 2-terminal transport measurements. Bottom: Schematic of the $MoS_2$ FET device. (b) Evolution of source-drain current, $I_{ds}$, with $V_g$ at 300K and $V_{ds} = 0.1$V for the device shown in (a). The red and black curves correspond to measurements before and after annealing respectively. Inset: same data on a semi-logarithmic scale.

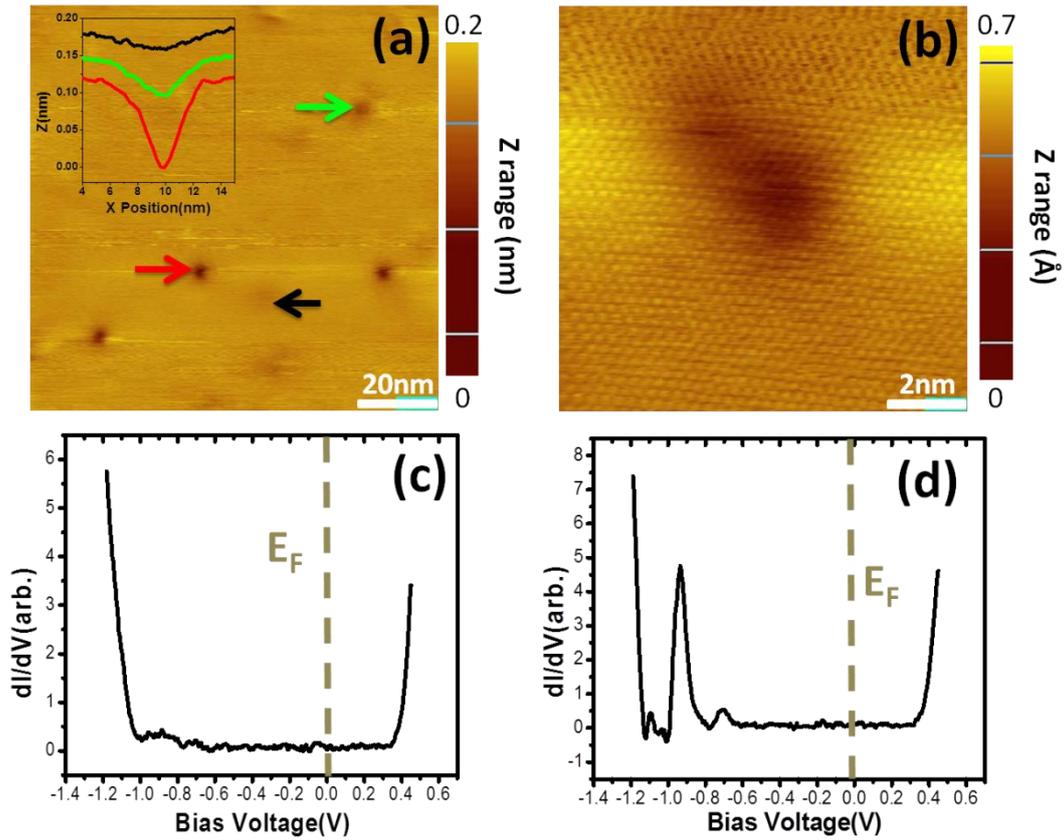

**Figure 4.** (a) STM constant current topography image of a 50nm × 50nm area at $V_b$ = 1.4V and I = 20 pA on bulk $MoS_2$ showing 3 types of defects with different apparent depths as indicated by the height profiles in the inset. (b) Constant current STM atomic resolution image on one of the defects. ($V_b$ = 1.2 V and I = 20 pA). (c) dI/dV spectrum taken far from any defect. (d) dI/dV spectrum taken on the defect shows a pronounced in-gap resonance near −0.94 V. The Fermi level is ~ 0.35 eV below the CB edge. The measurement was referenced to a set-point with tunneling resistance 15 GΩ (bias voltage $V_b$ =1.5V and tunneling current I =100 pA).

**RERERENCES**

# Supporting Information:

**Bandgap and doping effects in MoS$_2$ measured by Scanning Tunneling Microscopy and Spectroscopy**

*Chih-Pin Lu[1,2], Guohong Li[1], Jinhai Mao[1], Li-Min Wang[2] and Eva Y. Andrei[1]*

[1] Department of Physics and Astronomy, Rutgers University, Piscataway, New Jersey 08855, USA

[2] Department of Physics, National Taiwan University, Taipei 10617, Taiwan

**S1. Gated STS measurements.**

Special care has to be taken to carry out STS measurements when the Fermi energy lies in a gap. To avoid crashing the tip and damaging sample we used the following procedure:

1) The STM tip was retracted prior to any change in gate voltage, $V_g$.

2) Once $V_g$ was set a high set-point bias voltage was applied (>1.2V) to establish a stable tunneling junction. This was especially important for range of $V_g$ when the MoS2 enters the insulating regime.

3) Subsequently the set bias voltage should be reduced in order to resolve the bandgap. It is important to set the tunneling current as low as possible to minimize tip effects.

**S2. MOS$_2$ FET device performance**

The MoS$_2$ samples were annealed in forming gas (90% Ar, 10%H2) at 200 $^0$C for three hours prior to loading in the fridge. Before the measurement, the samples were further annealed in situ at 100 $^0$C and base pressure of ~ 10$^{-6}$ mbar for half an hour to remove water condensation. The results of the transport measurements are listed in the table below in the pre-anneal column. Following the first measurements the samples were further annealed at 110 $^0$C in high vacuum in situ for 20 hours, and then measured again. These results are listed in the post-anneal column. The mobility values for the devices are determined from $\mu = \frac{L}{WC_{ox}} \frac{1}{V_{ds}} \frac{dI_{ds}}{dV_g}$. The threshold voltage ($V_{gt}$) values are directly obtained from the $I_{ds}$ ($V_g$) curves as is the subthreashold swing (S).

We observed a systematic improvement in sample quality after *in-situ* annealing as measured by both mobility, μ and sub-threshold swing, *S*. Prior to annealing, both μ and *S* do not correlate with sample thickness. This situation changes after annealing. As shown in Figure S2, the post-anneal mobility displays a strong thickness dependence indicating that in this regime the mobility is dominated by surface scattering. At the same time, the absence of thickness scaling in the post-anneal values of *S* also indicates that it is dominated by surface traps. Our data thus suggests that annealing reduces bulk scattering and at the same time removes some of (but not all) the surface traps.

| Layer Number | pre-anneal | | | post-anneal | | |
|---|---|---|---|---|---|---|
| | Mobility (cm$^2$ V$^{-1}$ s$^{-1}$) | $V_{gt}$ (V) | S (V/dec) | Mobility (cm$^2$ V$^{-1}$ s$^{-1}$) | $V_{gt}$ (V) | S (V/dec) |
| ~15 layers | 2 | 10 | 1.5V/dec | 140 | -8 | 1.7V/dec |
| ~8 layers | 2 | 13 | 3V/dec | 50 | 16 | 2.6V/dec |
| 2 layers | 3 | 8 | 1.8V/dec | 6 | -10 | 0.7V/dec |
| 1 layer | 15 | 11 | 2.7V/dec | 18 | 0 | 1.9V/dec |

TableS1. Summary of parameters extracted from MoS$_2$ FET devices.

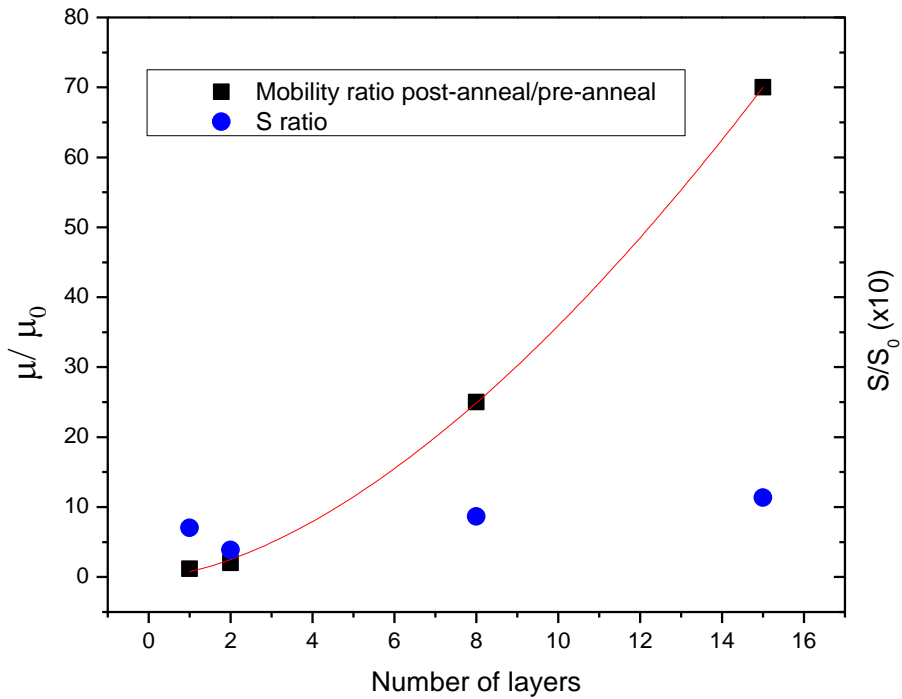

Figure S2. Ratio of the post-anneal to pre-anneal mobility and subthreshold swing as a function of sample thickness.

### S3. Atomic resolution topography of defects in 2H-MoS$_2$

Figure S3(a) shows an atomic resolution topography image of a freshly cleaved 2H-MoS$_2$ bulk sample. Figure S3(b) shows a height profile line cut taken along the dashed line in Figure S3(a). This gives an apparent depth of 0.05 nm modulated by 0.01 nm oscillation which correspond to individual atoms.

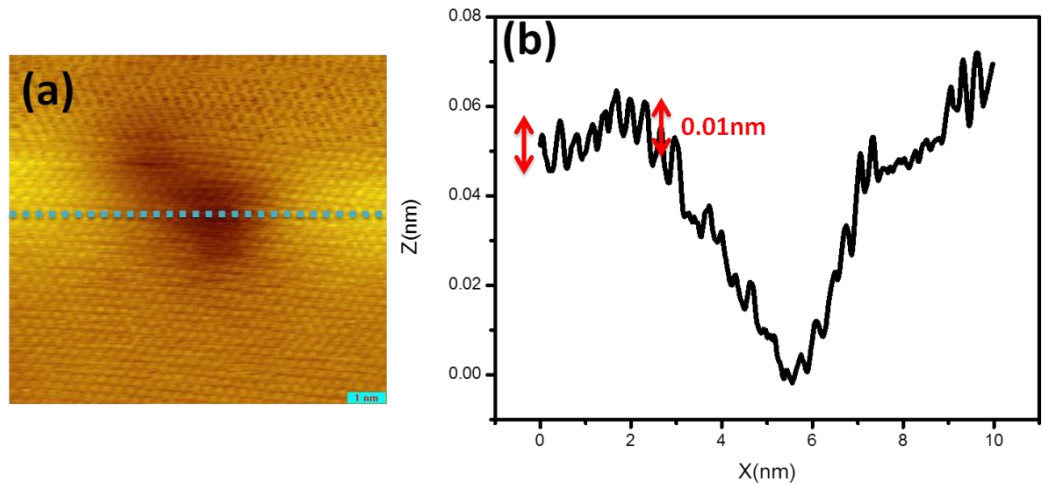

Figure S3. (a) STM constant current (I = 20pA) atomic resolution topography image in the vicinity of a defect seen on the surface of 2H-MoS$_2$ measured at $V_b$ = 1.2V. (b) Height profile taken along the dashed line in (a)

**S4. Topography of defects seen on the surface of bare and graphene covered 2H-MoS$_2$**

Figure S4 shows the STM topography images taken with different bias voltages. The images in Figure S4(a) and S4(b) were measured on the bare surface of bulk 2H-MoS$_2$ at $V_b$ = 1.5V and $V_b$ = −1.6V, respectively. The opposite sign of the contrast suggests that the feature represents a convolution of height and density of states. Repeating the measurement for graphene on a 15 layer MoS$_2$ sample deposited on a SiO$_2$ substrate shows similar defects. In this case too, the contrast on the defects changes sign with the polarity of the bias voltage, from dark spots at $V_b$ = 0.8V to bright at $V_b$ = −0.9V, as shown in Figure S4(c) and S4(d).

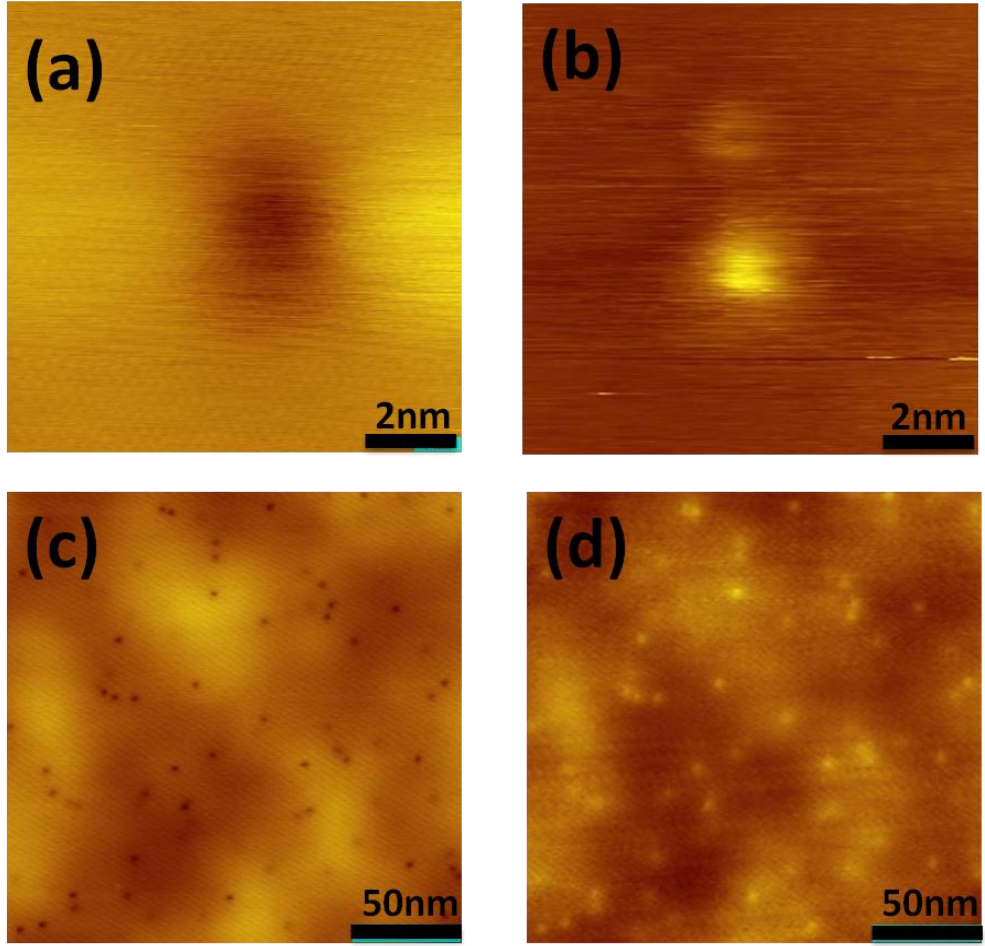

Figure S4. STM constant current (I = 20pA) topography image of defects on a freshly cleaved surface of bulk 2H-MoS$_2$. (a) Image of a 10nm × 10nm area measured at $V_b$ = 1.5V . (b Image of a 10nm × 10nm area measured at at $V_b$ = −1.6V. (c) Image of a 200nm × 200nm area of graphene on 2H-MoS$_2$ measured at $V_b$ = 0.8V. (d) Same as (c) with $V_b$ = −0.9V .

## 5. Concentration of defects

Figure S5 shows the STM topography image of a 200nm × 200nm area of freshly cleaved bulk 2H-MoS$_2$. To estimate the volume density of defects we measure the surface density of defects of equal size, $N_l \approx 3.5 \times 10^{10} cm^{-2}$ and assuming the same defect density in all S layers we obtain the volume density: $N_{3d} = N_l / a \approx 1.1 \times 10^{18} cm^{-3}$ where $a$ = 0.31 nm is the vertical separation between sulfur layers.

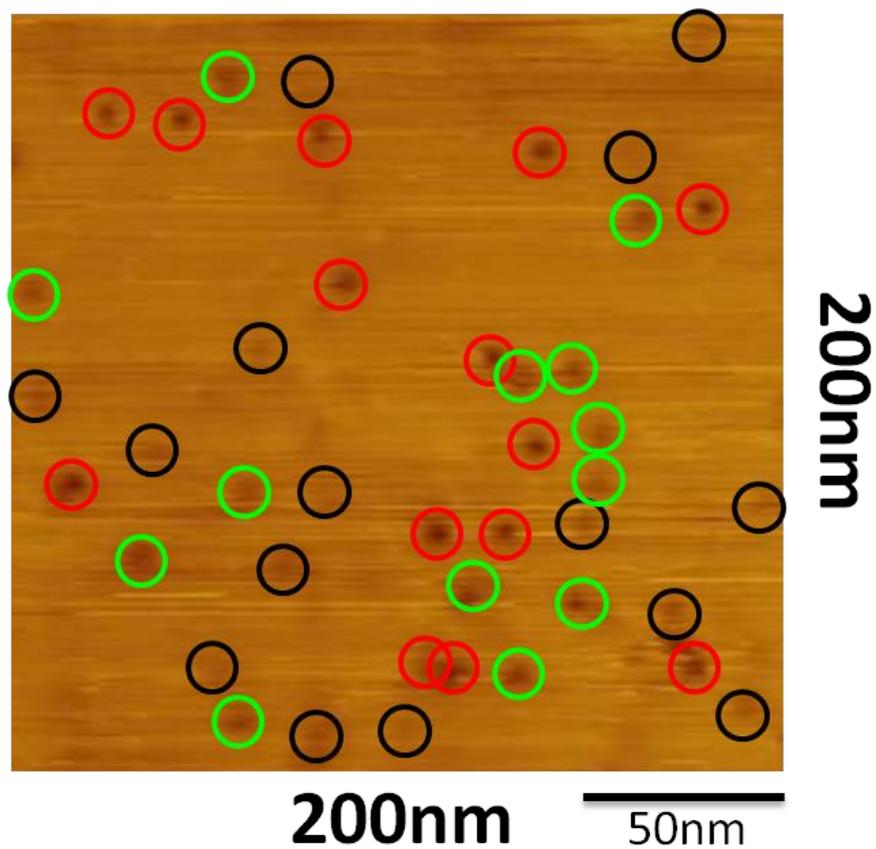

Figure S5. STM topography image of a 200nm × 200nm area taken at $V_b$ = 1.5V and I = 20pA on the freshly cleaved surface of bulk 2H-MoS$_2$ . The red, green and black circles correspond to apparent depths of ~ 0.12nm, 0.05nm and 0.02nm, respectively.